\documentclass[aps,prl,twocolumn,groupedaddress]{revtex4-1}

\usepackage{graphicx}
\usepackage{bm}
\begin{document}

\title{Accretion Dynamics on Wet Granular Materials}

\author{Guillaume Saingier}
\affiliation{Surface du Verre et Interfaces, UMR 125, CNRS/Saint-Gobain, 39, quai Lucien Lefranc, F-93303 Aubervilliers, Cedex France}
\author{Alban Sauret}
\affiliation{Surface du Verre et Interfaces, UMR 125, CNRS/Saint-Gobain, 39, quai Lucien Lefranc, F-93303 Aubervilliers, Cedex France}
\author{Pierre Jop}
\affiliation{Surface du Verre et Interfaces, UMR 125, CNRS/Saint-Gobain, 39, quai Lucien Lefranc, F-93303 Aubervilliers, Cedex France}

\begin{abstract}

Wet granular aggregates are common precursors of construction materials, food, and health care products. The physical mechanisms involved in the mixing of dry grains with a wet substrate are not well understood and difficult to control. Here, we study experimentally the accretion of dry grains on a wet granular substrate by measuring the growth dynamics of the wet aggregate. We show that this aggregate is fully saturated and its cohesion is ensured by the capillary depression at the air-liquid interface. The growth dynamics is controlled by the liquid fraction at the surface of the aggregate and exhibits two regimes. In the viscous regime, the growth dynamics is limited by the capillary-driven flow of liquid through the granular packing to the surface of the aggregate. In the capture regime, the capture probability depends on the availability of the liquid at the saturated interface, which is controlled by the hydrostatic depression in the material. We propose a model that rationalizes our observations and captures both dynamics based on the evolution of the capture probability with the hydrostatic depression.

\end{abstract}


\maketitle

Wet granular materials are common precursors of construction materials, food and health care products as well as relevant in many geophysical processes \cite{Herminghaus2005}. Indeed, the addition of liquid drastically modifies the behavior of a granular medium, and its rheological properties strongly depend on the proportion of the liquid \cite{Herminghaus2005, Nowak2005, Moller2007}. For large liquid volume fractions, a dense suspension is produced exhibiting fluidlike properties \cite{Bonnoit2010, Boyer2011}. By contrast, the presence of small amounts of liquid induces the formation of liquid bridges between grains, providing a strong cohesion to the material and a solidlike behavior \cite{Mason1999, Willett2000, Herminghaus2005, Mitarai2006, Kudrolli2008, Scheel2008}. These effects are commonly used in civil engineering processes that require mixing dry grains with a liquid to obtain new physical or chemical properties. Although the final product is homogeneous at the large scale, strong spatial heterogeneities in the liquid content are present during the blending, with rheological properties ranging from a dry state to a suspension in the mixture. Understanding how dry grains are incorporated into wet granular substrates thus requires coupling the dynamical interplay between the grains and the liquid. Most studies on the granulation in powders \cite{Iveson2001} consider the final size distribution of the aggregates. Other studies focus either on a static granular material, described as a porous medium, in contact with a fluid reservoir \cite{Delker1996, Reyssat2009, Xiao2012, Chopin2011}, or on the global rheological response during the blending process \cite{Cazacliu2009}. Only recently, some studies have considered the coupling between liquid and moving grains. For a low liquid content, the dry granular flow erodes the wet cohesive grains \cite{Lefebvre2013, Lefebvre2016}, whereas for a large liquid content, a stable and cohesive structure is built by accretion of dry grains on the wet granular phase \cite{Pacheco2012}. However, the local mechanisms and the accretion dynamics of dry grains onto a wet granular substrate are not well understood.

\begin{figure*}[t] 
\center
\includegraphics [height=53mm]{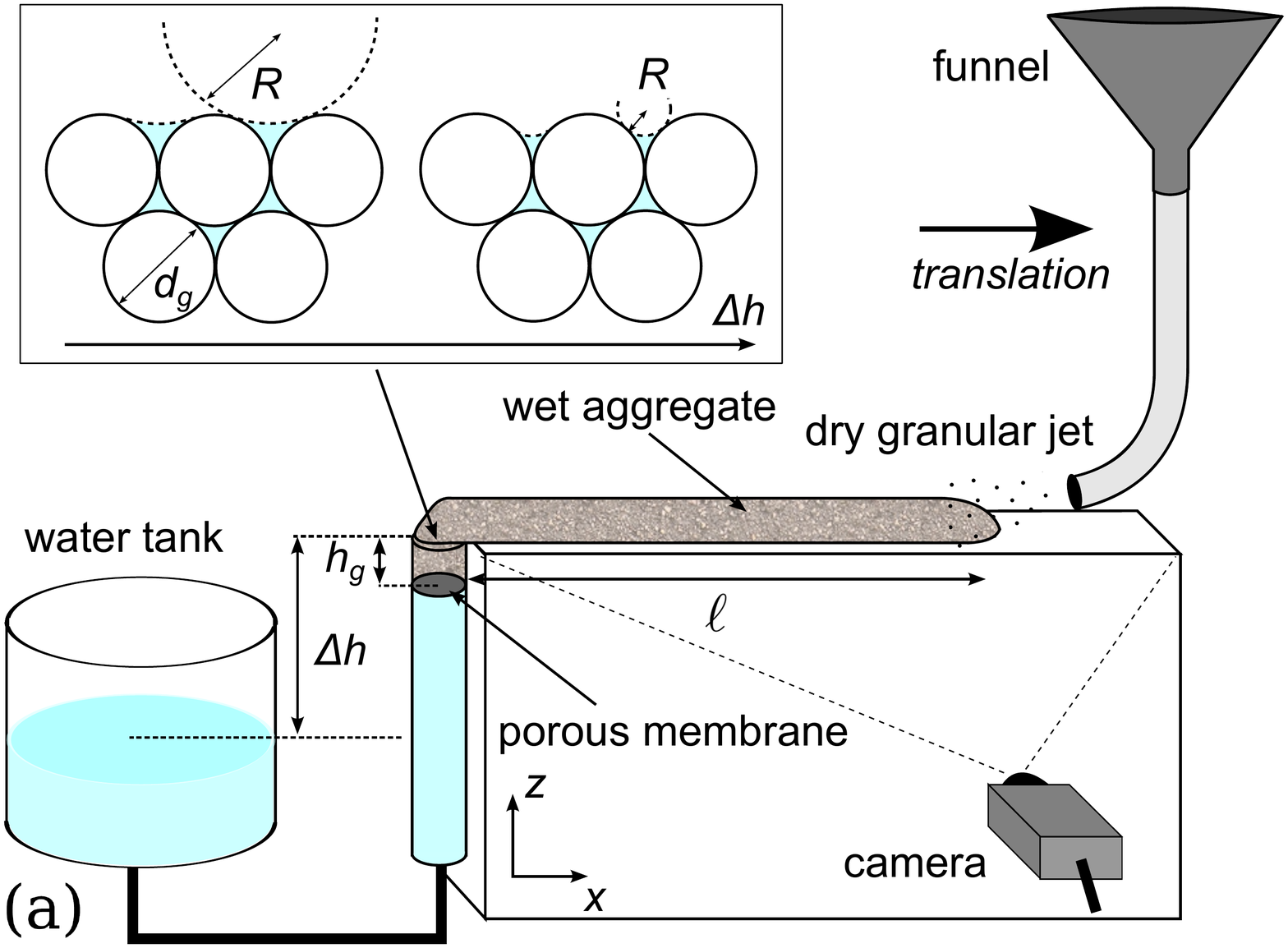}
\includegraphics [height=53mm]{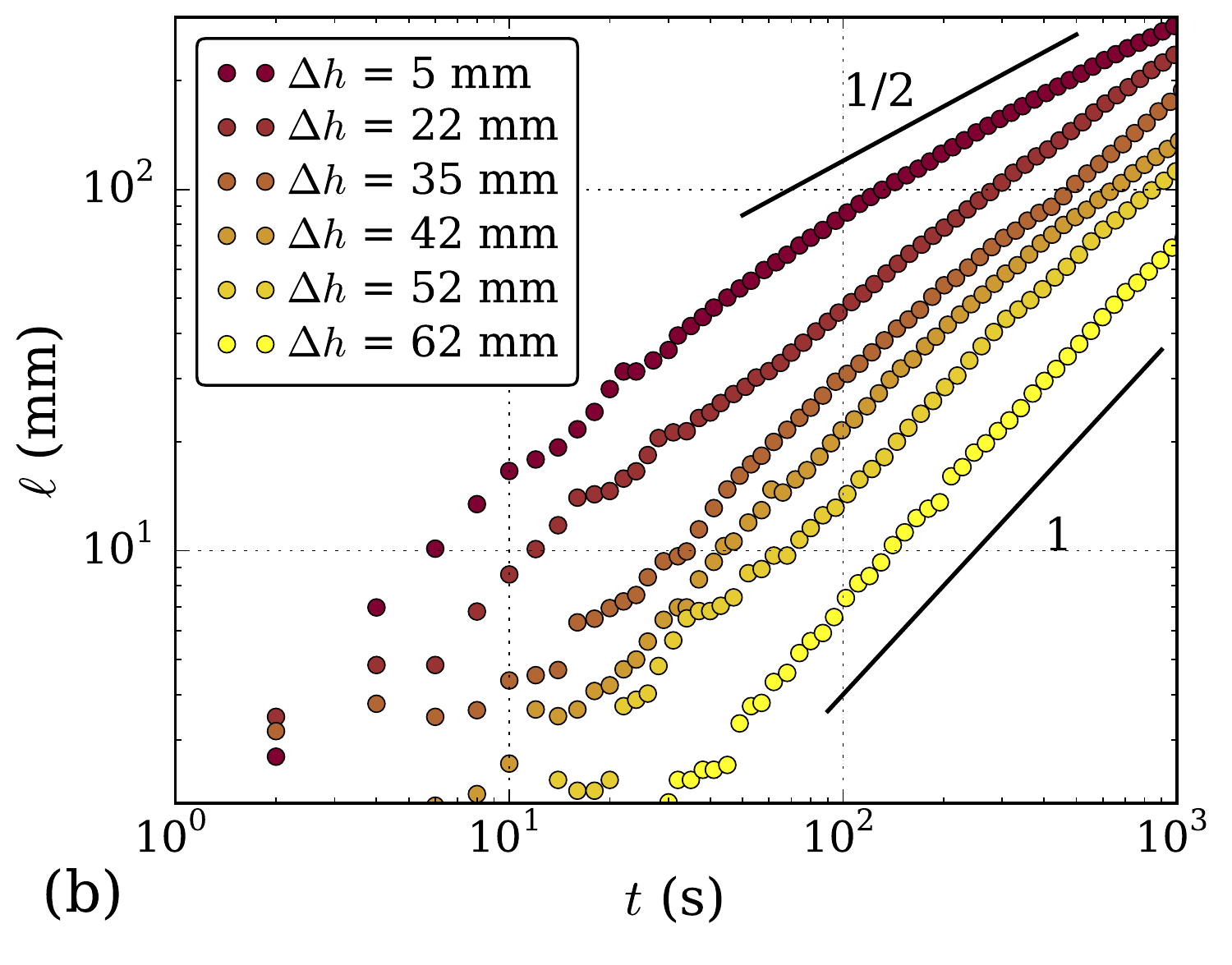}
\includegraphics [height=53mm]{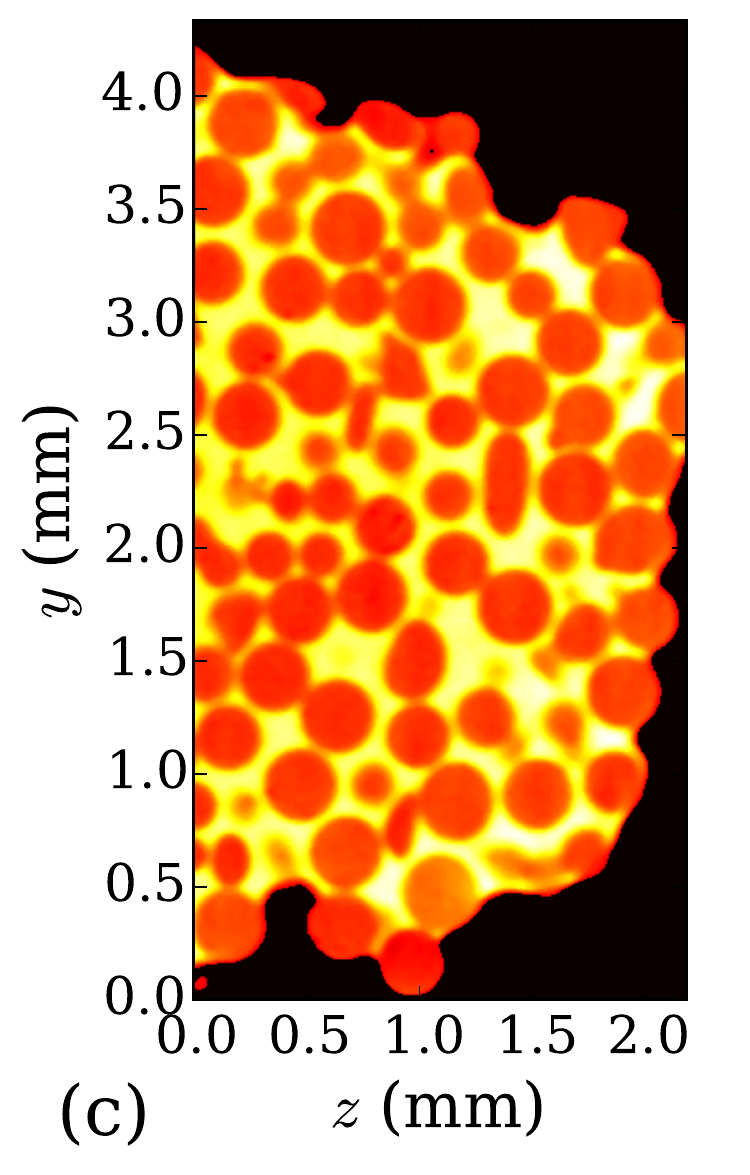}
\caption{(a) Schematic of the experimental set-up. Inset: Schematic of the meniscus for increasing $\Delta h$. (b) Time evolution of the length of the aggregate for different hydrostatic depressions, expressed as a function of $\Delta h$. (c) Cross section through a 3D tomogram of a wet aggregate. The liquid is colored in yellow whereas the glass beads are red and the air is in black.}
\label{set-up}
\end{figure*}

In this Letter, we investigate experimentally the accretion process between a static wet granular material and flowing grains using a model experiment presented schematically in Fig. 1. The accretion process results in the growth of the wet aggregate on a horizontal substrate. At the beginning of the experiment, the static wet granular substrate is made of spherical glass beads of diameter $d_g$ = 315-355 $\mu$m prepared in a vertical tube [Fig. \ref{set-up}(a)] with a height $h_g$ = 10 mm and a diameter equal to 10 mm. To keep the beads wet with a fixed hydrostatic pressure in the interstitial fluid, this substrate is connected to a water reservoir with an adjustable level through a porous membrane. We note $\Delta h$ the distance between the top of the substrate and the water level. The substrate is fully saturated, and the capillary pressure drop associated to the local curvature of the menisci balances the hydrostatic depression at the liquid- air interface:
\begin{equation}
p_0 - \frac{2 \gamma \cos\theta}{R} = p_0 - \rho g \Delta h,
\label{pressure_balance}
\end{equation} 
where $p_0$ is the atmospheric pressure, $\gamma$ = 71 mN/m ($\pm$ 1 mN/m) the surface tension of the water, $\rho$ = 1 g/cm$^3$ the water density, $\theta$ = 27$^\circ$ ($\pm$ 5$^\circ$) the contact angle of the water on a glass bead \cite{Raux2013} and $1/R$ the mean curvature of the meniscus at the interface. 

To study the accretion, identical dry glass beads are poured on the wet substrate at constant flow rate $Q_g$ = 1.1 g/s using an inox funnel of diameter 3 mm connected to a 20-cm-long and 6 mm diameter inox tubing with a flexible end. A grid is placed between the funnel and the tube to disperse the grains and create a diluted jet collimated by the tube. The grains are ejected with an angle of about 45$^\circ$ with the horizontal at constant velocity ($v_g$ $\simeq$ 1.6 m/s) controlled, in the diluted regime, by the length of the tube. As the aggregate grows by the accretion of dry grains, the funnel is moved away at the growth velocity, so that the grains are released at a constant distance from the wet substrate, typically 5 mm. Note that the accretion occurs on a horizontal plane, so the hydrostatic depression in the interstitial liquid remains constant during an experiment. A thin PMMA plate of width 6 mm is used to support the weight of the aggregate during its growth and avoid the accumulation of grains that are not trapped, which bounce off after the impact. PMMA is chosen as it is less hydrophilic than the glass beads and does not influence the growth dynamics \cite{Suppl}. The growth of the aggregate is recorded at 0.5 Hz with a CCD camera and analyzed by image processing.

%
We investigate the role of the hydrostatic depression by performing systematic experiments at different water heights in the tank, $\Delta h$. The growth dynamics is reported in Fig. \ref{set-up}(b), where we plot the length of the aggregate $\ell$ as a function of time for a constant $Q_g$ and $v_g$. The growth rate decreases when $\Delta h$ increases, indicating that the accretion process is less efficient for a large hydrostatic depression. Moreover, the dynamics drastically evolves with $\Delta h$ and exhibits a smooth transition from a diffusive regime at low $\Delta h$, where $\ell$ is proportional to $t^{1/2}$, and a linear regime at large $\Delta h$, where $\ell$ is proportional to $t$ [Fig. \ref{set-up}(b)]. To understand how the liquid is distributed in the aggregate during the accretion process, we image the microstructure with X-ray tomography \cite{Suppl, Scheel2008}.
The 3D reconstruction shows that the aggregate is fully saturated without any air bubble for any value of $\Delta h$ [Fig. \ref{set-up}(c)]. The aggregate is in a capillary state and the cohesion of the structure results from the capillary depression at the air/liquid interface \cite{Mitarai2006}. Therefore, during an experiment, the curvatures of the menisci are in equilibrium with the local pressure along the wet aggregate. 
%

%
%
\begin{figure*}[t] 
\center
\includegraphics [height=45.0mm]{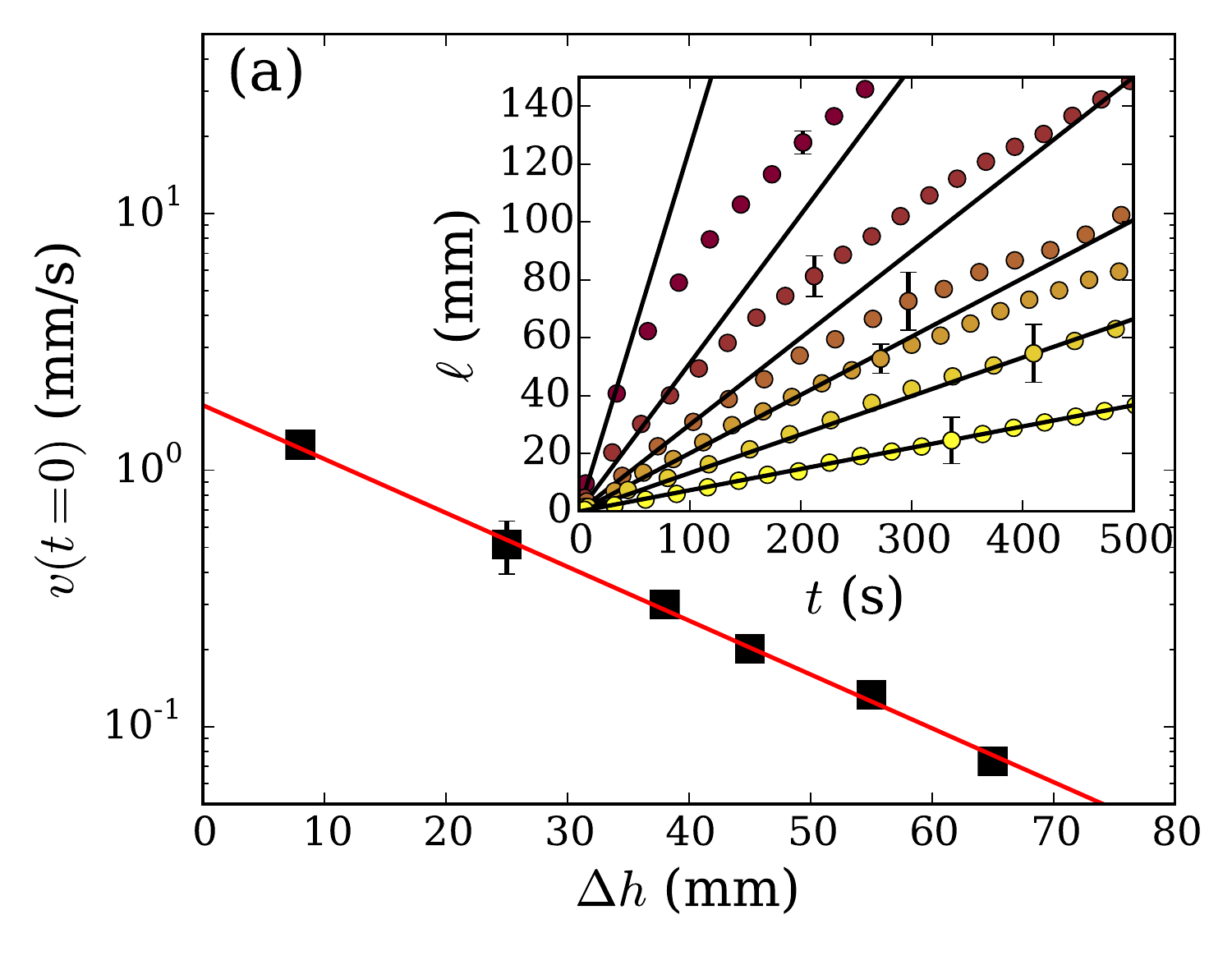}
\includegraphics [width=58mm]{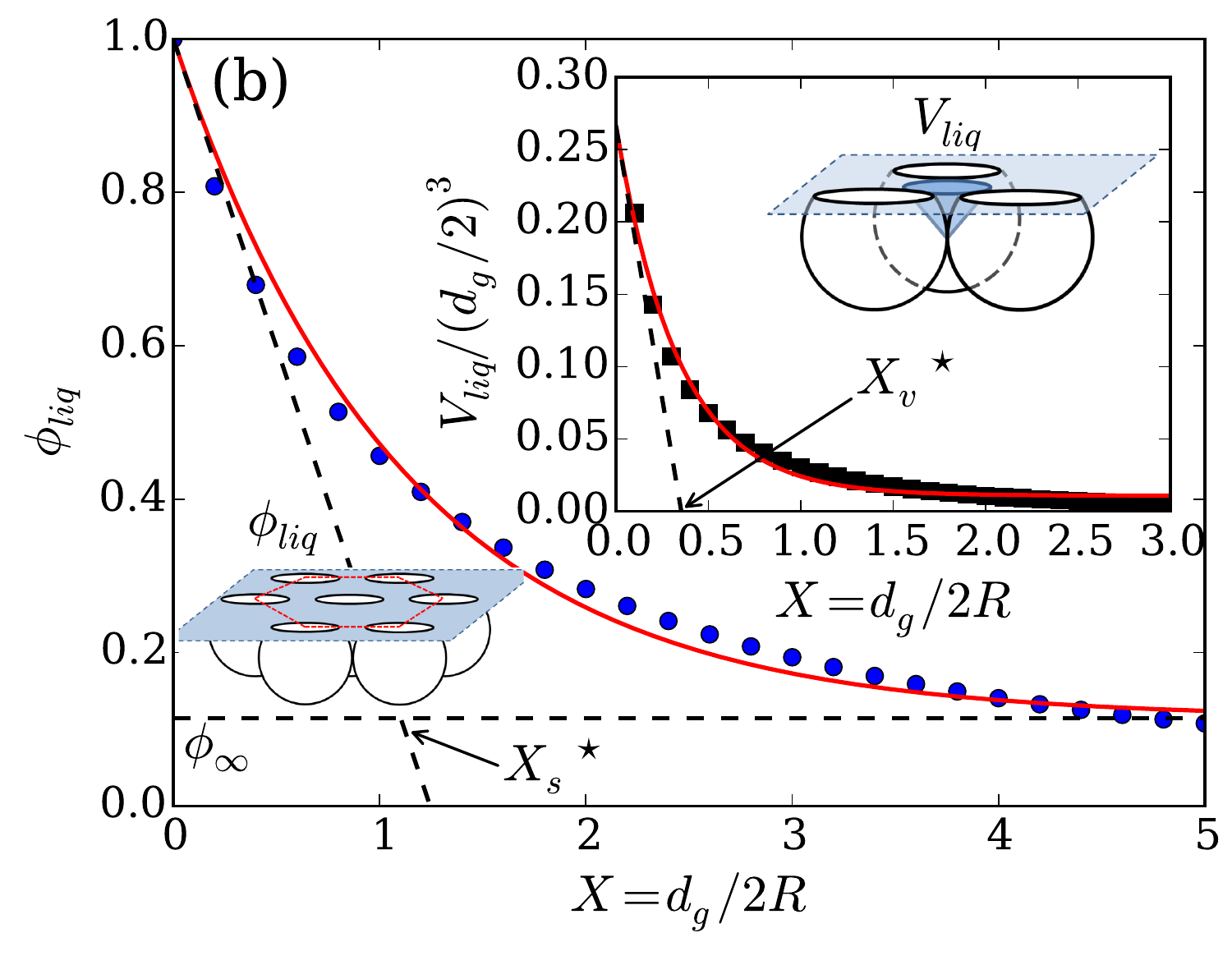}
\includegraphics [width=58mm]{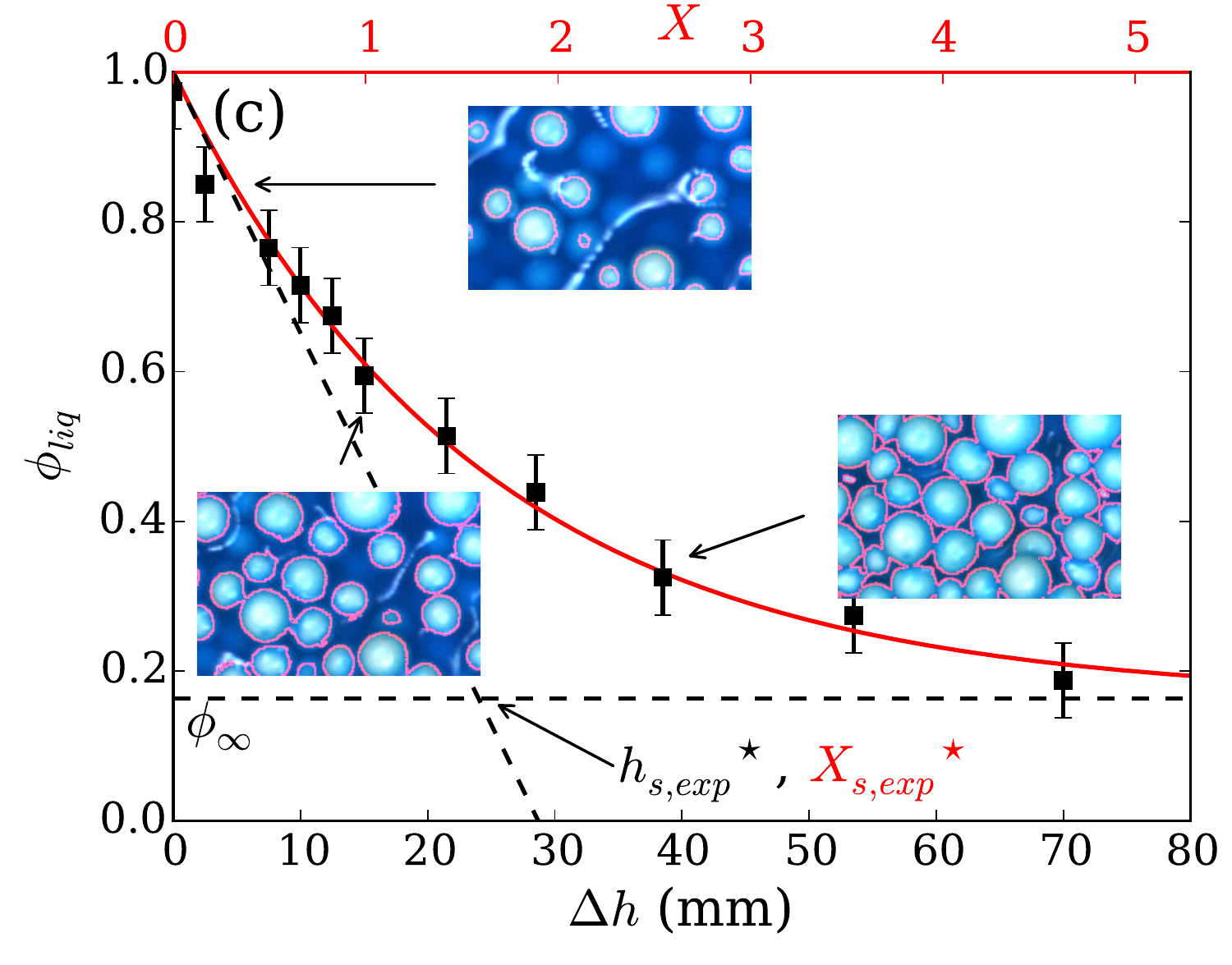}
\caption{(a) Initial growth rate of the aggregate $v(t=0)$ as a function of the height $\Delta h$. The fit gives $h^\star$ = 20 mm ($X^\star$ = 0.26). Inset: Temporal evolution of the aggregate length. (b) Evolution of the liquid fraction $\phi_{liq}$ at the liquid/air interface in a wet granular packing as the function of the ratio $X = 2 d_g/R$. Data are fitted by $\phi_{liq} = \phi_\infty + (1-\phi_\infty)\exp(-X/{X_s}^\star)$  with ${X_s}^\star$ = 1.10. Inset: Evolution of the dimensionless volume associated to a conical site of capture - ${X_v}^\star$ = 0.34. (c) Experimental evolution of the apparent liquid surface as function of height $\Delta h$. Images are obtained by binocular microscopy with an opaque dyed liquid and zirconium beads of 500 $\mu$m diameter. The red line is an exponential expression with ${h_{s,exp}}^\star$ = 24 mm $\pm$ 4 mm corresponding to ${X_{s,exp}}^\star$ = 1.59 $\pm$ 0.26. Experimental parameters: $\gamma$ = 44 mN/m ($\pm$ 2 mN/m), $\theta$ = 65$^\circ$ ($\pm$ 5$^\circ$)]}.
\label{sticking_proba}
\end{figure*}
To understand the existence of these two different regimes, we propose a local growth mechanism by granular accretion. The aggregate growth is directly related to the fraction of dry grains captured at the liquid interface of the wet material. As dry grains are added, the liquid has to penetrate into the granular packing to reach its equilibrium position and to be accessible to the impacting granular jet. At low hydrostatic depression (small $\Delta h$) [inset of Fig. \ref{set-up}(a)], the air/liquid interface is slightly curved and easily available to capture a large fraction of impacting grains. In this case, the growth dynamics is limited by the viscous displacement of the fluid into the granular packing. This viscous regime is modeled by the Darcy's law connecting the flow velocity to the driving pressure corresponding here to the capillary pressure in the pores. The Lucas-Washburn equation describes the imbibition in all the granular structure of total length $L^{v}$, which is the sum of the aggregate length $\ell^{v}$ and the substrate length $h_g$ \cite{Washburn1921}:
\begin{equation}
L^{v}(t) = \ell^{v}(t) + h_g = \sqrt{\frac{2 k \Delta p}{\eta}(t + t_0)},
\label{viscous_regime}
\end{equation}
where $k$ is the permeability of the packing, $\eta$ is the dynamic viscosity of the fluid, $t_0$ is the time for the liquid to penetrate into the substrate and $\Delta p = p_c - \rho g \Delta h$ the capillary pressure reduced by the hydrostatic depression. The pressure $p_c$ is associated to the pore radius $r_p$ and defined as $2\gamma\cos\theta/r_p$. In the following, $r_p$ is taken equal to the grain radius $d_g/2$ \cite{Reyssat2009}. Note that $t_0$ is equal to ${h_g}^2 \eta/2 k \Delta p^{h_g}$, where $\Delta p^{h_g} = p_c - \rho g (\Delta h -h_g)$. In the range of height $\Delta h$ investigated, $t_0$ varies between 0.7~s and 3~s and will be neglected in the following as it remains small compared to the time scale of our experiments. In this regime, the aggregate length scales as $t^{1/2}$ in agreement with the experimental results.

A second regime is explored at large $\Delta h$ corresponding to large hydrostatic depressions. In this situation the menisci are strongly deformed and the liquid is less accessible to the impacting grains. The growth dynamics of the aggregate is then limited by the efficiency of the capture process. Consequently, the growth rate can be defined using the fraction of grains captured over the amount of grains impacting the aggregate. Introducing the capture probability $\mathcal{P}_{capt}$ and assuming that this probability is constant during an experiment, where $\Delta h$ remains constant, the growth dynamics in the capture regime is:
\begin{equation}
\ell^{c}(t) = \frac{Q_g}{\rho_s \phi S}\,\mathcal{P}_{capt} \, t,
\label{capture_regime}
\end{equation}
where $\rho_s$ = 2.5 g/cm$^3$ is the grain density, $\phi$ = 0.63 $\pm$ 0.01 is the compacity of the aggregate and S its cross-section.

The transition from the viscous regime to the capture regime occurs when the typical growth rates associated to {those} two limiting mechanisms are comparable. Equating the {growth rates} leads to a typical time scale $t_c$ and a typical length scale $\ell_c$ characterizing the growth process. These parameters depend on the capture probability and are defined as:
\begin{equation}
t_c = \frac{k}{2\eta} \left( \frac{\rho_g \phi S}{\mathcal{P}_{capt} Q_g} \right)^2 \Delta p, \ \ \ \ell_c = \frac{k}{2\eta} \frac{\rho_g \phi S}{\mathcal{P}_{capt} Q_g}  \Delta p.
\label{non_scaled}
\end{equation}
Moreover, our experiments show that the capture probability $\mathcal{P}_{capt}$ decreases with the height $\Delta h$, thus with the hydrostatic depression. To estimate this variation, we compute the initial growth velocity $v(t=0)$ for each experiment. At the beginning of the growth, the accretion process is not limited by the rate of imbibition through the short porous aggregate, but only by the capture rate of the first grains. As shown in Fig. \ref{sticking_proba}(a), the initial growth rate decreases exponentially with the altitude $\Delta h$ and the variation of the capture probability reads: 
\begin{equation}
v(\Delta h, t=0) = v_0 \exp\left( - \frac{\Delta h}{h^\star} \right) = \frac{Q_g}{\rho_s \phi S}\mathcal{P}_{capt}(\Delta h),
\end{equation}
{thus,}
\begin{equation}
\mathcal{P}_{capt}(\Delta h) = \mathcal{P}_0 \exp \left( - \frac{\Delta h}{h^\star} \right), 
\end{equation}
where $h^\star$ = 20 mm is the length characterizing the {velocity} decrease and $\mathcal{P}_0$ = $(\rho_s \phi S v_0)/Q_g$ is the capture probability at the water level  with $v_0$ = 1.86 mm/s. A similar expression was found empirically for the rising velocity of a vertical granular tower \cite{Pacheco2012}. {We defined the dimensionless curvature $X$ = $d_g/2R$ which is related to the height $\Delta h$ by Eq. (\ref{pressure_balance}) such that $X = (d_g \rho g / \gamma \cos \theta) \Delta h$, which gives the dimensionless length associated to the velocity decrease, $X^\star$ = $(d_g \rho g / \gamma \cos \theta) \,h^\star$ = 0.26.}

\begin{figure}[t] 
\center
\includegraphics [height=55mm]{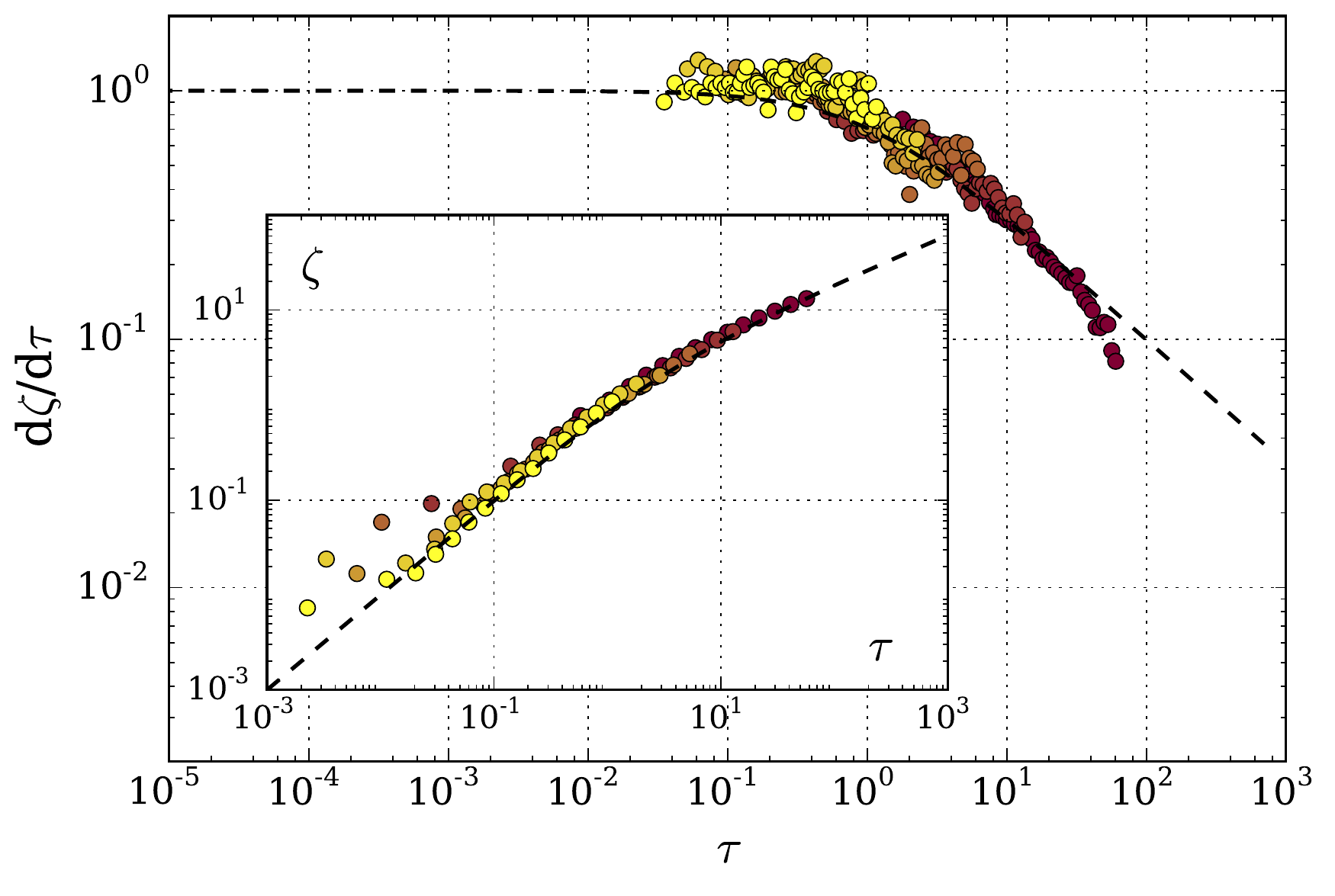}
\caption{Evolution of the rescaled growth velocity $\rm{d}\zeta/\rm{d}\tau$ as a function of the rescaled time $\tau$ defined in Eq. \ref{non_scaled} for all hydrostatic depressions. Inset: Evolution of the rescaled length of aggregate $\zeta$ as the function of the rescaled time $\tau$. The dashed lines correspond to the theoretical predictions given by Eq. (\ref{sol_theo_position}).}
\label{results_resc}
\end{figure}

To explain the expression of the probability, we propose a crude geometrical model that relates on the liquid distribution at the air/liquid interface with the probability to capture a grain. As the aggregate is fully saturated, the hydrostatic depression, associated with $\Delta h$, retracts the liquid menisci. Thus the interfacial liquid area and volume decrease. We first approximate the air/liquid interface by a plane located at the bottom of a spherical meniscus and intersecting a dense layer of spheres (see the schematics of Fig. \ref{sticking_proba}(b) and \cite{Suppl} for details of the calculation). Furthermore, we assume that the liquid is perfectly wetting the beads which are organized in a hexagonal lattice \cite{Suppl}. Under these assumptions, we calculate the liquid area as a function of $X$ [Fig \ref{sticking_proba}(b)]. In addition, we report the liquid volume calculated using the cone inscribed between the spheres [inset of Fig \ref{sticking_proba}(b)] \cite{Suppl}. Both the evolution of the area and the volume are fitted by an exponential decay in agreement with the expression proposed for the capture probability. We define ${X_s}^\star$ and ${X_v}^\star$ as the characteristic dimensionless length associated to each decay, respectively [Fig. \ref{sticking_proba}(b)]. Fitting the numerical data leads to ${X_s}^\star$ = 1.10 and ${X_v}^\star$ = 0.34. The value of ${X_v}^\star$, compared to $X^\star$, suggests that the capture probability depends on the volume of liquid available between the interfacial grains. Furthermore, the evolution of the liquid distribution is obtained by direct imaging of the liquid at the interface of a wet granular material for varying $\Delta h$ [see Fig. \ref{sticking_proba}(c) and \cite{Suppl} for experimental details]. An exponential decrease is observed with ${X_{s,exp}}^\star$ = 1.59 $\pm$ 0.26, consistent with the computed value. 
Using the values of $v_0$ and $h^\star$ determined experimentally, we compute the length $\ell_c$ and the time $t_c$ and  introduce the rescaled parameters $\zeta$ = $\ell/\ell_c$ and $\tau$ = $t/t_c$. The rescaled data are plotted in Fig. \ref{results_resc} for different heights {\cite{Suppl}}. The data collapse well on a master curve, confirming that the aggregate growth is dominated by the competition between the sticking properties of the grains and the properties of the flow in a porous medium. 
Our model based on the menisci retraction as well as our tomographic reconstructions show that air bubbles are not present in the aggregate as suggested by Pacheco \textit{et al.} who claim that granular towers can emerge both in funicular and capillary states. Their argument was developed to explain their measurements of an apparent exponential decrease of the ratio between the mass of liquid and the mass of grains along a vertical granular tower \cite{Pacheco2012}. Our results demonstrates that the accretion is an interfacial phenomenon, which only takes place in the capillary state. Finally, our model also provides an explanation for the logarithmic rise dynamics recorded for vertical towers, which takes place in the capture regime.
%

%
To predict the full accretion dynamics and the smooth transition between the two regimes, we introduce two characteristic times. Indeed, to increase the length $\ell$ of the aggregate by one grain diameter $d_g$, we can separate the process into a capture time $\tau_{capt}$ associated to the capture process followed by a viscous time $\tau_{visc}$ corresponding to the fluid motion in the last layer of grains until the equilibrium position of the meniscus is reached. These times are estimated using the growth rate associated to each phenomenon (see Eqs. \ref{viscous_regime} and \ref{capture_regime}):
\begin{equation}
\tau_{capt}  = \frac{d_g \rho_g \phi S}{Q_g \mathcal{P}_{capt}}, \ \ \ \tau_{visc} =  \frac{\eta}{k}\frac{\ell d_g}{\Delta p}.
\end{equation}
Summing these expressions, we obtain the aggregate growth rate $v = d_g/\delta t$. The dimensionless equations of the length as a function of the time during the accretion are:
\begin{equation}
\frac{\rm{d} \zeta}{\rm{d} \tau} = \frac{1}{1 + \zeta/2}, \ \ \ \text{and thus} \ \ \
\label{sol_theo_velo}
\zeta(\tau) = 2(\sqrt{1+\tau}-1).
\label{sol_theo_position}
\end{equation}
These predictions are compared to our experimental measurements in Fig. \ref{results_resc}. The smooth transition from the capture regime to the viscous regime is well captured by our model, which highlights the coupling between the fluid dynamics and the grains motion.

We now discuss the trapping mechanisms. The initial kinetic energy $E_i$ of one impacting grain must be dissipated during the capture. Three main mechanisms have been identified. First, Crassous \textit{et al.} have shown that the restitution coefficient of a grain bouncing on a dry granular pile ranges between $0.3$ and $0.5$ for our inclination (30$^\circ$ - 60$^\circ$), which represents an energy loss from 75\% to 90\% of $E_i$ \cite{Crassous2007}. Also, several works studied the capture of a grain by a flat liquid film \cite{Davis2002, Antonyuk2010, Gollwitzer2012, Muller2016}, and showed the role of the viscous dissipation and the rupture distance of capillary bridges to predict the sticking condition. Here, based on a typical rupture distance $d_g/4$, the viscous and capillary energy dissipations are of the order of 5\% and 10\% of $E_i$ {\cite{Suppl}}. Assuming that these three phenomena are independent, we conclude that the initial kinetic energy may be completely dissipated. However, the calculation of the probability $\mathcal{P}_0$ from the experimental value of $v_0$ indicates that only 2.5\% of the grains are trapped at $\Delta h=0$, which means that only 2.5\% of the aggregate interface is able to capture a grain. This low efficiency can be explained by the crucial role of the position of the grain impact to fully dissipate the kinetic energy. If the liquid depth is too small, the grain will bounce off.

In conclusion, the flow of a dry granular material on a wet granular substrate induces an accretion process characterized by the growth of the saturated phase by the accretion of grains. We show that this capture is a local process controlled by the capture probability of grains, which is related to the liquid availability at the interface. The horizontal accretion process reveals two distinct regimes, depending on the mechanism that limits the presence of fluid at the surface of the aggregate, either the viscous displacement in the porous material or the hydrostatic depression. We propose a theoretical model that predicts the correct transition and dynamics. This study provides a solid grounding to understand the interaction between flowing granular media and a fluid flow.

\begin{acknowledgments}
We are grateful to William Woelffel for his help and his advice for the tomographic acquisitions and we acknowledge support from Saint-Gobain Recherche to access their lab tomograph. This work benefitted from the financial support of French ANRT (PhD. CIFRE 2015/0504).
\end{acknowledgments}


%

\end{document}